\documentclass{solarphysics}

\usepackage[optionalrh,showbiblabels]{spr-sola-addons} 
\usepackage{graphicx}        
\usepackage{color}           
\usepackage{url}

\newcommand{\mateja}[1]{\textcolor{black}{#1}}

\chardef\us=`\_

\begin{document}

\begin{article}
\begin{opening}

\title{The catalog of Hvar Observatory solar observations}

\author[addressref=aff1,corref,email={mdumbovic@geof.hr}]{\inits{M.}\fnm{Mateja}~\lnm{Dumbovi\'c}}
\author[addressref=aff1]{\inits{L.}\fnm{Lu\v{c}i}~\lnm{Karbonini}}
\author[addressref=aff1]{\inits{J.}\fnm{Ja\v{s}a}~\lnm{\v{C}alogovi\'c}}
\author[addressref=aff1]{\inits{F.}\fnm{Filip}~\lnm{Matkovi\'c}}
\author[addressref=aff1]{\inits{K.}\fnm{Karmen}~\lnm{Martini\'c}}
\author[addressref=aff1]{\inits{A.K.}\fnm{Akshay Kumar}~\lnm{Remeshan}}
\author[addressref=aff1]{\inits{R.}\fnm{Roman}~\lnm{Braj\v{s}a}}
\author[addressref=aff1]{\inits{B.}\fnm{Bojan}~\lnm{Vr\v{s}nak}}

\address[id=aff1]{Hvar Observatory, Faculty of Geodesy, University of Zagreb}

\runningauthor{Dumbovi\'c et al.}
\runningtitle{HVAR Catalog}
\begin{abstract}

\mateja{We compile the catalog of Hvar Observatory solar observations in the time period corresponding to regular digitally stored chromospheric and photospheric observations 2010-2019. We make basic characterisation of observed phenomena and compare them to catalogs which are based on full disc solar images. We compile a catalog of observed ARs consisting of 1100 entries, where each AR is classified according to McIntosh and Mt Wilson classifications. We find that HVAR observations are biased towards more frequently observing more complex ARs and observing them in longer time periods, likely related to the small FOV not encompassing the whole solar disc. In H$\alpha$ observations we catalog conspicuous filaments/prominences and flares. We characterise filaments according to their location, chirality (if possible) and eruptive signatures. Analysis of the eruptive filaments reveals a slight bias in HVAR catalog towards observation of partial eruptions, possibly related to the observers tendency to observe filament which already showed some activity. In the flare catalog we focus on their observed eruptive signatures (loops or ribbons) and their shape. In addition, we associate them to GOES soft X-ray flares to determine their corresponding class. We find that HVAR observations seem biased towards more frequently observing stronger flares and observing them in longer time periods. We demonstrate the feasibility of the catalog on a case study of the flare detected on 2 August 2011 in HVAR H$\alpha$ observations and related Sun-to-Earth phenomena. Through flare-CME-ICME association we demonstrate the agreement of remote and in situ properties. The data used for this study, as well as the catalog, are made publicly available.}

\end{abstract}
\end{opening}
\section{Introduction}
\label{intro} 

Though solar observations today are dominantly relying on satellites or large, high-resolution ground-based observatories, small traditional solar observatories still produce science quality data that can supplement state-of-the art observations from modern instruments. For this reason we have decided to compile the catalog of Hvar Observatory solar observations.

The solar telescope was installed at Hvar Observatory in 1972 \citep[for a recent review of the history of Hvar Observatory see][]{brajsa17} and consists of a photospheric and chromospheric telescopes, that have parallel optical axes and are mounted as one unit on the same German parallax mounting \citep{ambroz77, calogovic12}. The double solar telescope thus observes simultaneously photosphere and chromosphere, in white light (WL) and in H$\alpha$ line, respectively. The corresponding fields of view are smaller than the full solar disk (about 7 and 11 arcmin), i.e. typically the size of the active region. Therefore, it is very important to specify the location of the observed feature and its classification. Due to the atmosphere, the best resolution is $\approx1$ arcsec, and the instruments are capable of achieving this resolution under favorable atmospheric conditions. However during the usual weather conditions (turbulences, wind, thin clouds, humidity), the telescope rarely achieves this resolution. 

\mateja{Given their characteristics (as described in more detail in Section \ref{instruments}) the instruments are traditionally considered as most suitable for studying fast active processes on larger scales \citep{ambroz82}. These include H$\alpha$ flares \citep[e.g.][]{ruzdjak89}, quiescent prominences \citep[e.g.][]{ruzdjak91}, prominence oscillations \citep[e.g.][]{vrsnak93}, and eruptive prominences \citep[e.g.][]{vrsnak88,vrsnak90,vrsnak91} and Moreton waves \citep{Moreton60,vrsnak08}. We note, however, that in the whole time span of analysed HVAR  H$\alpha$  observations we did not observe a single Moreton wave. HVAR WL observations are suitable for analysis of individual sunspot groups. For instance, HVAR WL observations were found to be suitable for an investigation of proper motions of sunspot groups \citet{wohl03,wohl04}. With a FOV encompassing the size of the typical AR, HVAR WL observations are suitable to be used for study of individual properties of active regions, as their complexity and morphology is tightly related to the (flaring) activity \citep[e.g.][and references therein]{toriumi19}. Finally, on short time scales we might expect to observe white-light flares \citep[see e.g.][]{hudson06b,benz17}. We note however, that in the whole time span of analysed HVAR WL observations we did not observe a single WL flare. Therefore, in the catalog we focus on active regions, flares and filaments/prominences.}

Throughout its lifetime the telescope was mostly used in scheduled observing campaigns and during observing visits. The regular chromospheric and photospheric observations with Hvar Observatory (HVAR) double solar telescope using modern equipment began in May 2011 after the installation, adjustment, and testing of the new equipment \citep{calogovic12}. These continued until 2019, when the regular observations were hampered and became rarer due to technical reasons. Therefore, in this first version of the catalog we encompass all observations performed by the double solar telescope at Hvar Observatory in the time period May 2011 -- November 2019. The data in this period are stored in the archive\footnote{\url{http://oh.geof.unizg.hr}}, which is publicly accessible together with the observer logs.

\mateja{In this study we go beyond the observers log and make basic characterisation of observed phenomena that might be of scientific use for modern research. The paper is organised as follows: in Section \ref{instruments} we briefly describe the instruments and observations for the catalog, in Section \ref{catalog} we describe what we focus on in the catalog and how it is built, along with providing access to the catalog, and finally in Section \ref{case_study} we demonstrate the usefulness of the catalog in a case study combining multispacecraft and multiwavelength observations.}

\section{Description of Hvar solar H$\alpha$ and white light telescopes}
\label{instruments}

\mateja{The double solar telescope at Hvar Observatory consists of two Carl Zeiss refractors for H$\alpha$ and white light (WL) observations.
The main purpose of the Hvar solar telescope was to perform regular photospheric and chromospheric observations with optimal angular and temporal resolution following the observing routines of the Ondrejov Observatory \citep{ambroz77}. Later, a similar imaging system with CCD cameras as at the Kanzelh\"ohe Solar Observatory (KSO) was installed to complement the KSO full-disk observations with high-resolution Hvar images of active regions \citep{Otruba-2005, Otruba-2008, calogovic12}. In 2010, a new generation of CCD cameras (Pulnix TM-4200GE 12-bit) was installed on the telescopes, allowing a larger dynamic range (very bright or faint phenomena, e.g. flares and prominences) as well as higher spatial resolution and time cadence \citep{calogovic12}.}    

\mateja{The photospheric (WL) telescope has a main objective with a diameter of 217 mm and a focal length of 2450 mm. It consists of a Baader AstroSolar filter and an iris diaphragm that both reduce the intensity of the sunlight in the telescope. In addition, solar continuum and UV/IR blocking filters are used to increase the contrast of sunspots and solar granulation by allowing the spectral range around 540 nm to pass through, free of emission and absorption lines. Such a setup has a corresponding field of view of about 11 arcmin, which results in a theoretical resolution of 0.33 arcsec/pixel when using a 4Mpix Pulnix CCD (2048x2048 pixels). However, this resolution cannot be achieved due to the ideal theoretical diffraction limit of 0.6 arcsec with an objective of 217 mm and an observing wavelength band of 540 nm. Such a resolution is also never achieved, since the atmosphere also has a considerable influence on the resolution. Under the best seeing conditions, the resolution can be around 1 arc second. In the system used, the images are therefore oversampled, which helps to reduce aliasing and noise through image processing and thus improves the image resolution.}

\begin{table}
\caption{H$\alpha$ and WL observation information and statistics for the Hvar Solar Telescope in the period from May 2011 to November 2019.}
\label{tab_hst_stats}
\centering
\small
\begin{tabular}{lcc} 
\hline
 &	H$\alpha$ observations	&	WL observations \\
\hline
Field of view & 7 arcmin & 11 arcmin \\
Cadence	& 15 sec & 60 sec \\
Number of observing days & 735 & 710 \\
Number of images & 694 000 & 175 000 \\
Total duration & 2891 h (120.49 days) & 2927 h (121.96 days) \\
Size in archive & 5.35 Tb & 1.40 Tb \\
\hline
\end{tabular}
\end{table}

\mateja{Chrompsheric (H$\alpha$) telescope with a main objective diameter of 130 mm consists of an energy reduction filter, an adjustable iris diaphragm, an additional teleconverter lens, and an H$\alpha$ filter. A solar spectrum research grade filter with a passband of 0.2 {\AA} is used as the H$\alpha$ filter. The corresponding field of view of such a system is around 7 arcmin, which results in 0.21 arcsec/pixel resolution with the 4Mpix CCD used. The theoretical diffraction limit of the lens with a diameter of 130 mm and the wavelength band of 656 nm is 1.3 arcsec. Similar to the WL telescope, the images are therefore oversampled.}

\mateja{The acquisition software almost identical to the KSO \citep{Otruba-2005}, regulates the exposure time automatically and carries out the image selection in real time. The software automatically selects the sharpest image from about 10 images taken by the CCD within about 1 second to avoid the blur caused by the turbulence and to select the moments with good seeing. In normal observation mode, a FITS image is saved every minute for photospheric observations and every 15 seconds for chromospheric observations (see Table \ref{tab_hst_stats}).}

\mateja{Some information and statistics about H$\alpha$ and WL observations can be found in Table \ref{tab_hst_stats}. In total, there are more than 700 days of continuous observations with a total duration of almost 3000 hours in the period from May 2011 to November 2019. Figure \ref{tab_hst_stats} shows the distribution of all individual observations during this period. Due to weather conditions, the frequency of observations is denser in summer than in winter, with some gaps in 2015 and 2019 when technical problems hampered observations.}

\begin{figure}
\centerline{\includegraphics[width=0.95\textwidth]{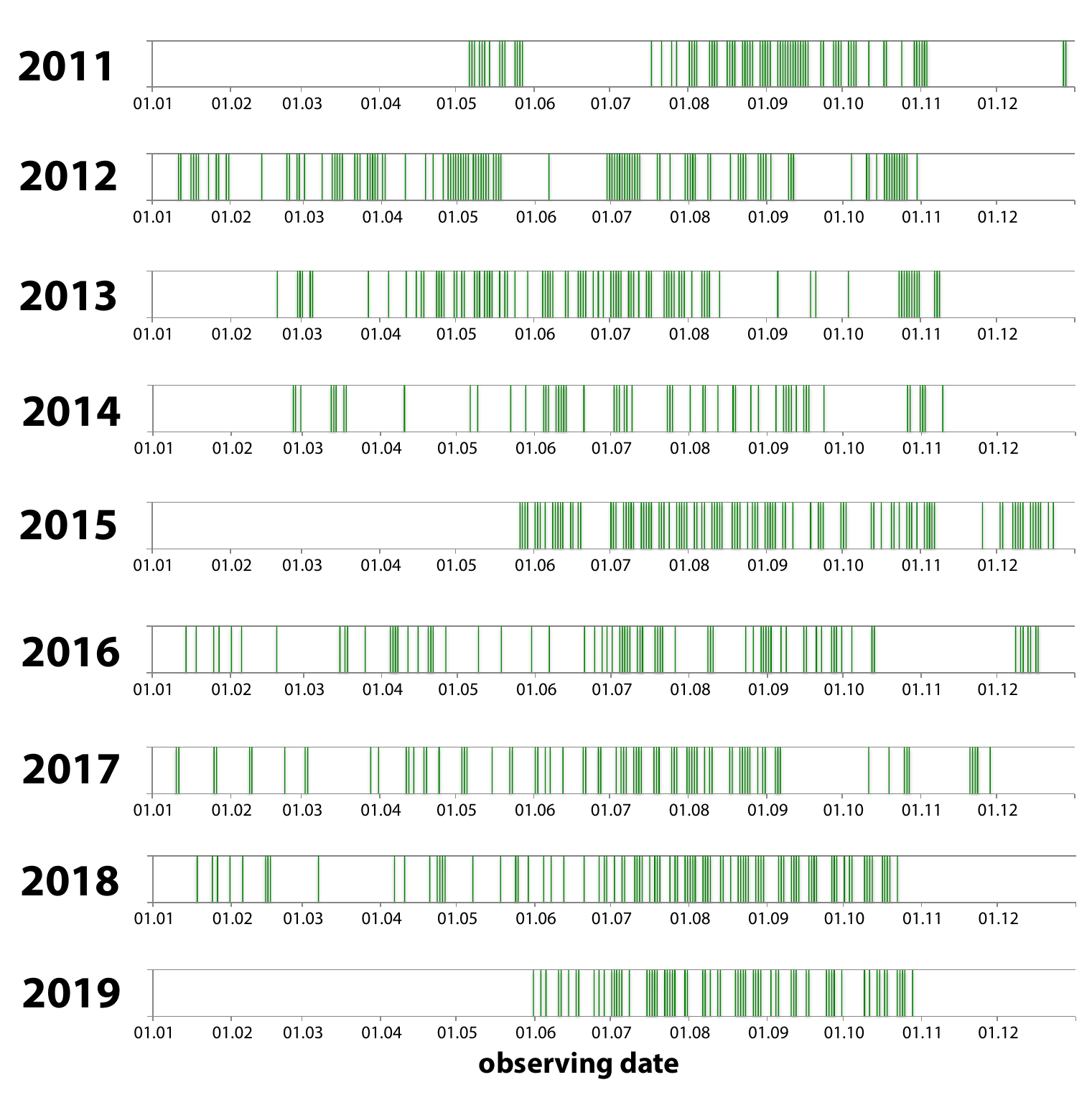}}
\caption{ Distribution of individual H$\alpha$ and WL observations in the period from May 2011 to November 2019.}
\label{fig_hst-stats}
\end{figure}

\mateja{The movies of the daily observations in this time period are publicly available at Hvar Observatory web page\footnote{\url{http://oh.geof.unizg.hr}}, whereas the data (in FITS or jpeg format) is available upon request. The basic observation information, such as which regions were observed and what were the weather conditions, are provided in the observers log, also available at the same web page.}

\section{Catalog of HVAR WL and H$\alpha$ observations}
\label{catalog}

\subsection{White light observations of the solar photosphere} 
\label{WL}

\begin{figure}
\centerline{\includegraphics[width=0.95\textwidth]{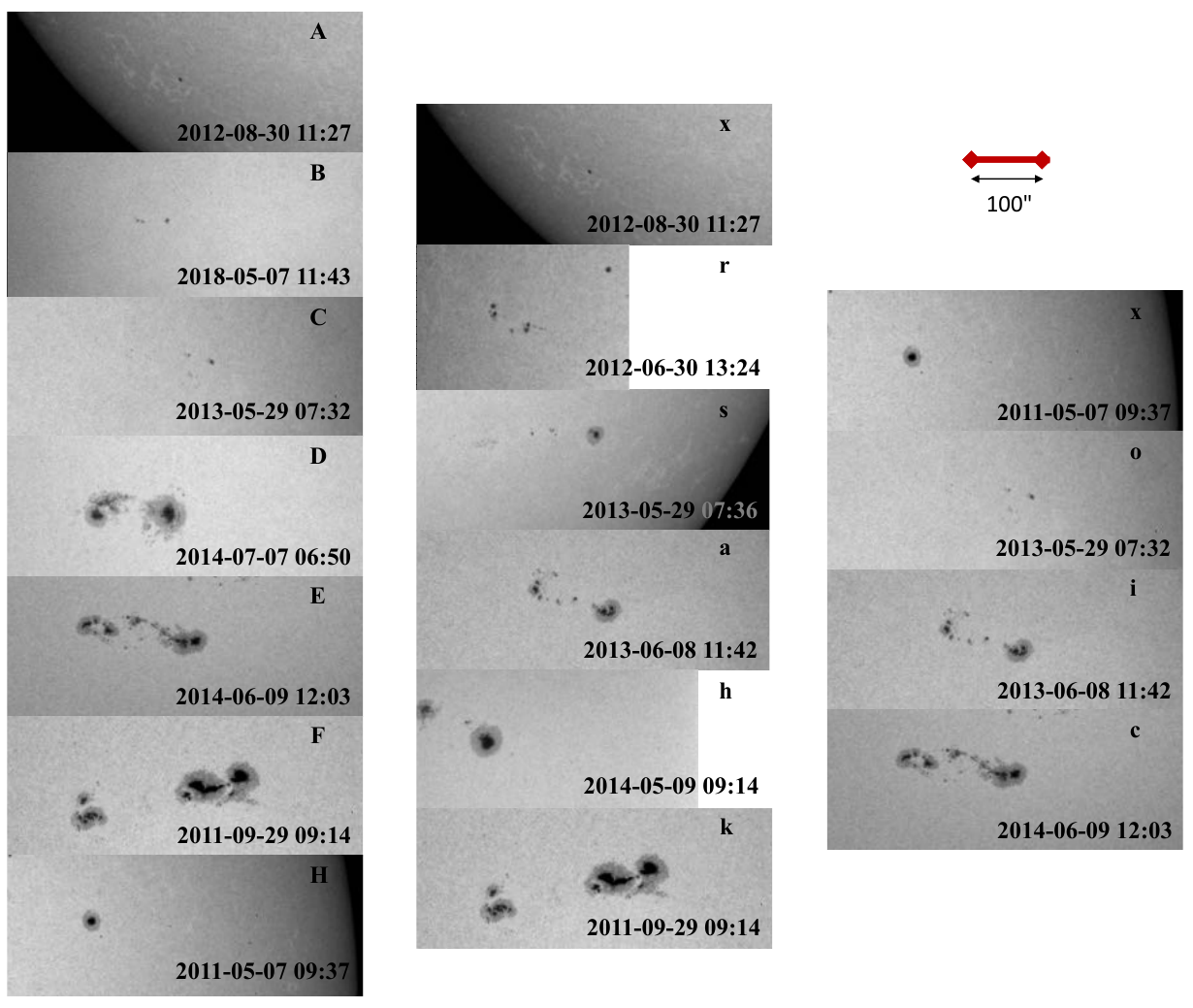}}
\caption{Selected examples from the HVAR WL observations to describe the general form of the McIntosh classification. 
\textbf{The left panels} show 7 classes of the modified Zurich classification system: Class A -- unipolar, no penumbrae; Class B -- bipolar, no penumbrae; Class C -- bipolar, penumbra on one end of the group; Class D -- bipolar, penumbra on spots at both ends of the group, length $<10^{\circ}$; Class E -- bipolar, penumbra on spots at both ends of the group, $10^{\circ}<$length$<15^{\circ}$; Class F -- bipolar, penumbra on spots at both ends of the group, length$>15^{\circ}$; Class H -- unipolar, penumbra.
\textbf{The middle panels} show 6 classes of the penumbra of the largest sunspot: Class x -- no penumbra; r -- rudimentary penumbra; s -- small, symmetric penumbra; a -- small, asymmetric penumbra; h -- large, symmetric penumbra; k --large, asymmetric penumbra.
\textbf{The right panels} show 4 classes of the sunspot distribution:   Class x -- undefined relative spottedness (for unipolar groups); Class o -- open spottedness (few and small interior spots); Class i -- intermediate spottedness (numerous interior spots, but without penumbra); Class c -- compact spottedness (numerous interior spots, some with penumbra)}
\label{fig_class1}
\end{figure}

To build HVAR WL observations catalog we focus on the classification of the active regions observed. One of the early categorisations used for sunspot groups is the Zurich sunspot classification, which attempts to describe the evolution of the sunspot group to increasingly larger structures. This categorisation was later modified by \citet{mcintosh90} into a form used nowadays. The modified Zurich sunspot classification takes into account not only the size of the group, but also whether or not the penumbra is present and how it is distributed and consists of 7 classes \citep[for details see][]{mcintosh90}. The modified Zurich sunspot classification is supplemented by the classification of the properties of the penumbra of the largest sunspot and sunspot distribution into a joint McIntosh classification \citep{mcintosh90}. The general form of this classification is Zpc, where Z is the modified Zurich class, p is the class of the penumbra of the largest sunspot, and c describes the sunspot distribution \citep[for details see][]{mcintosh90}\footnote{see also the SWx classification website: https://www.stce.be/educational/classification}. 
There are 60 possible Zpc combinations that make McIntosh classification made by combinations of 7 Z-classes, 6 p-classes, and 4 c-classes (note that not all Z, p, and c classes can be combined with one another). Different Z, p, and c classes are shown in Figure \ref{fig_class1} on selected examples from the HVAR WL observations.

\begin{figure}
\centerline{\includegraphics[width=0.95\textwidth]{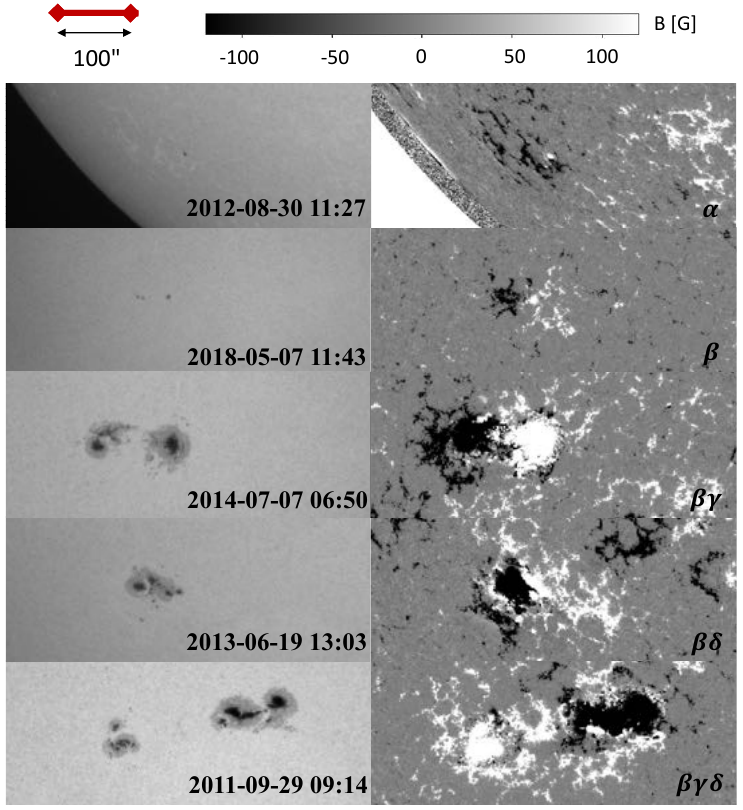}}
\caption{Selected examples from the HVAR WL observations (left panels) alongside HMI observations (right panels) to describe the general form of the Mt Wilson classification. Top-to-bottom panels show: Class $\alpha$ -- unipolar; Class $\beta$ -- simple bipolar; Class $\beta\gamma$ -- complex bipolar; Class $\beta\delta$ -- Class $\beta$ with $\delta$ spot; Class $\beta\gamma\delta$ -- Class $\beta\gamma$ with $\delta$ spot. $\delta$ region marks a region of closely placed opposite polarity umbrae within a common penumbra.
}
\label{fig_class2}
\end{figure}

Next, we classify observed active regions according to the Mount Wilson classification, which refers to the magnetic structures of active regions (ARs). For that purpose we use the observations from the Solar Dynamics Observatory \citep[SDO][]{pesnell12} Helioseismic and Magnetic Imager \citep[HMI][]{schou12} available through JHelioviewer service \citep{muller17}. Mt Wilson classification scheme is based on classes devised by \citet{hale19} describing 4 basic magnetic types and 3 more complex classes derived from simple Hale classes added by \citet{kunzel65}. Out of 7 different Mt Wilson classes, 5 are shown in Figure \ref{fig_class2} on selected examples from the HVAR WL observations alongside HMI observations (we do not find examples of classes $\gamma$ and $\gamma-\delta$ in HVAR WL observations).

\mateja{The active regions were inspected visually by the observer (L.K.) using movies of the daily observations available at Hvar Observatory web page. Since HVAR data lacks information on the relative distance from the solar center, we did not determine the spatial extent of the active region using HVAR data, but instead we compared HVAR data with full disc WL images available at SolarMonitor\footnote{https://www.solarmonitor.org} and JHelioviewer \citep{muller17}, where the spatial extent was measured by the observer using the grid of stonyhurst longitude and latitude and the polarities were examined visually using SDO/HMI observations. Based on the visual inspection and the determination of the spatial extent the observer determined the classes of the active regions using criteria described by Figures \ref{fig_class1} and \ref{fig_class2}.} In addition, we cross-check our observations with the full-disc photospheric observations available at SolarMonitor\footnote{https://www.solarmonitor.org}. In our catalog we provide NOAA number of the AR based on the visual comparison of HVAR WL and SolarMonitor observations for a given day. The SolarMonitor provides McIntosh and Mt Wilson classifications for each observed AR which in most cases agrees with the one we derived from HVAR WL observations, but not always. This is likely related to our inability to compare HVAR WL observations at an exact time with SolarMonitor daily image based on which the AR class was derived.

In addition to classifying the observed ARs, we also provide a measure of the quality of the observation, as determined by the observer, the quality index, QI. QI ranges from 1 (worst) to 5 (best). Note that QI is a somewhat subjective measure, as it encompass both the ``true'' image/data quality (due to e.g. weather and atmospheric conditions) as well as the observers uncertainty in categorising the event. Nevertheless, we find that the QI might be a useful information for future users of the catalog, especially given that it was consistently assigned by the same observer.

\begin{table}
\caption{The statistics of the AR classes in the HVAR WL catalog according to McIntosh and Mt Wilson classification}
\label{tab_ARs}
\centering
\small
\begin{tabular}{cccccccc} 
\hline
McIntosh class	&		&		&		&		&		&		&		\\
\hline
	            &	A	&	B	&	C	&	D	&	E	&	F	&	H	\\
	            &	59	&	89	&	234	&	307	&	169	&	56	&	186	\\
\hline
	            &	x	&	r	&	s	&	a	&	h	&	k	&		\\
	            &	148	&	95	&	301	&	306	&	47	&	203	&		\\
\hline
	            &	x	&	o	&	i	&	c	&		&		&		\\
	            &	245	&	525	&	173	&	157	&		&		&		\\
\hline
Mt Wilson class	&		&		&		&		&		&		&		\\
\hline
	            &	$\alpha$	&	$\beta$	&	$\beta-\gamma$	&	$\gamma$	&	$\beta-\delta$	&	$\beta-\gamma-\delta$	&	$\gamma-\delta$	\\
	            &	244	&	375	&	280	&	0	&	26	&	175	&	0	\\
\hline
\end{tabular}
\end{table}

We observe and catalog a total of 1100 ARs, most of which are of moderate complexity: C or D modified Zurich class, with small symmetric or asymmetric penumbra present and of open spottedness. Regarding the Mt Wilson classification, more than 50\% of all ARs are of either $\beta$ or $\beta-\gamma$ class. The full statistics of the AR classes observed is given in Table \ref{tab_ARs}. The catalog of ARs observed in HVAR WL is publicly available through figshare\footnote{\url{https://doi.org/10.6084/m9.figshare.23256890.v1}}.

\begin{figure}
\centerline{\includegraphics[width=0.95\textwidth]{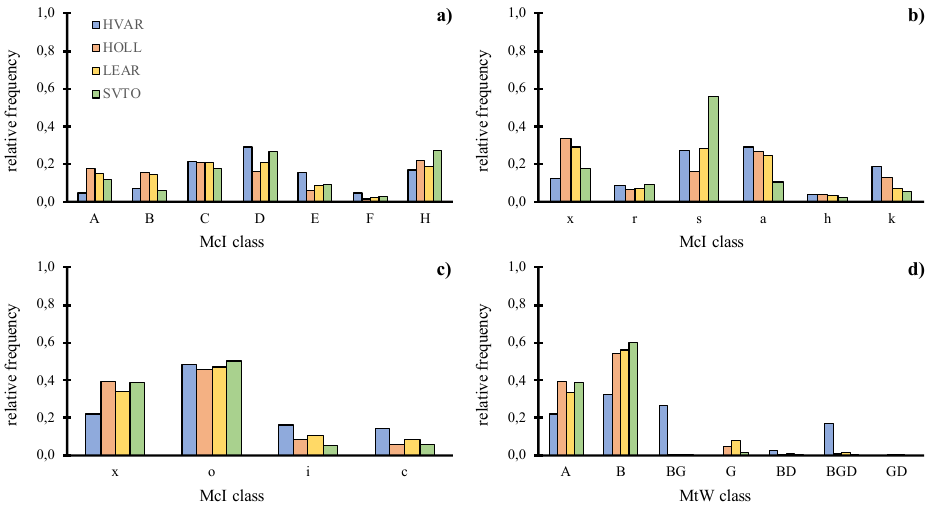}}
\caption{Relative frequencies of observed AR classes cataloged by HVAR, HOLL, LEAR and SVTO observatories in the time period from May 2011 until July 2017. a-c correspond to the modified Zurich classes and d corresponds to the Mt Wilson classification.
}
\label{fig_WL_stats1}
\end{figure}

\mateja{We compare HVAR WL AR observation and classification to those cataloged by the U.S. Air Force (USAF) solar region reports from three observatories Holloman (HOLL), Learmonth (LEAR) and San Vito dei Normanni (SVTO) available through National Geophysical Data Center (NGDC) of the National Oceanic and atmospheric administration (NOAA)\footnote{\url{https://www.ngdc.noaa.gov/stp/space-weather/solar-data/solar-features/sunspot-regions/usaf_mwl/}}, available in the time period up to July 2017. We thus consider time period from May 2011 until July 2017. In this time period HVAR catalogs 1017 ARs, HOLL 8671, LEAR 9233, and SVTO 8681. In Figure \ref{fig_WL_stats1} we compare relative frequencies of different classes cataloged by 4 observatories in the aforementioned time period. We can see that for modified Zurich classes there are no notable differences between HVAR and other observatories. There is however a notable difference in the Mt Wilson classification. We can see that while other observatories mostly have ARs classified as alpha and beta, there is a significant proportion of ARs classified as $\beta-\gamma$ and $\beta-\gamma-\delta$ in HVAR catalog. This could be related to a possible HVAR bias towards selecting and tracking more complex and flare-rich ARs to observe due to small FOV.}

\begin{figure}
\centerline{\includegraphics[width=0.95\textwidth]{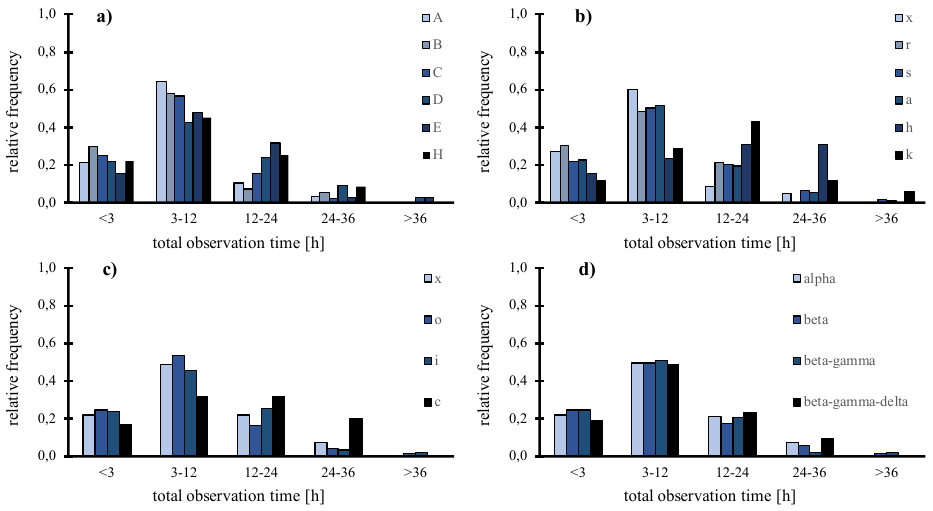}}
\caption{Relative frequencies of the total observing time dedicated to ARs of different AR classes. a-c correspond to the modified Zurich classes and d corresponds to the Mt Wilson classification. Note that in the d-panel we omitted classes which were not observed.}
\label{fig_WL_stats2}
\end{figure}

\mateja{In order to check this potential bias we analyse total observing time dedicated to a specific AR of specific class. Note that a particular AR can be observed on multiple occasions and the class of the AR can change during the total observing time. Thus, for the purpose of this analysis we associate a particular AR with a class that was appointed to it at the time of first observation. Figure \ref{fig_WL_stats2} shows relative frequencies of the total observing time dedicated to ARs of different AR classes. It can be seen that there is a slight tendency that more complex ARs are tracked for longer periods.}

\subsection{H$\alpha$ observations of the solar chromosphere} 
\label{Halpha}

In building the HVAR H$\alpha$ catalog we focus on observations of prominences, filaments, and solar flares. We focus only on conspicuous prominences/filaments and flares and mark the duration of their observation in the HVAR data archive as well as some basic properties. As ``conspicuous'' we consider prominent features that are either well observed, large, and/or very active and interesting/unusual. We note that this introduces a certain bias of the observer, nevertheless our motivation was to constrain the catalog to features which 1) we could characterise by more than sheer time of observation and 2) \mateja{might be interesting to compare with associated coronal mass ejections and ultimately their interplanetary counterparts}. Furthermore, each feature is marked by a quality index, QI, similarly as with ARs described in Section \ref{WL}. We note that in the case of active and eruptive phenomena such as prominence oscillations, filament eruptions, and flares additional factor influencing the QI is the duration of observation (e.g. whether the observations include all stages of the phenomena from the pre-activity stage to the post-activity stage).

The prominences/filaments are first characterised according to their location and whether we observe them as bright (prominence) or dark feature (filament). Based on where they are located, we differ quiescent, AR, and intermediate filaments. Quiescent filaments are most common, they are usually found well away from ARs, extend to large heights, and can typically last for two to three solar rotations. AR filaments are relatively long and narrow and are usually too low to appear as prominences at the limb, whereas intermediate filaments are  found at the borders of AR. Intermediate and AR filaments are located at mid-latitudes, while quiescent filaments may exist over all latitudes on the Sun \citep{foukal,mackay10}. For active and intermediate filaments/prominences we identify relevant AR (marked by NOAA AR number in the catalog) by comparison with SolarMonitor or JHelioviewer full disc images in H$\alpha$ when possible, otherwise using SDO Atmospheric Imaging Assembly \citep[AIA][]{lemen12} 304 \AA\, wavelength observations through JHelioviewer.

\begin{figure}
\centerline{\includegraphics[width=0.95\textwidth]{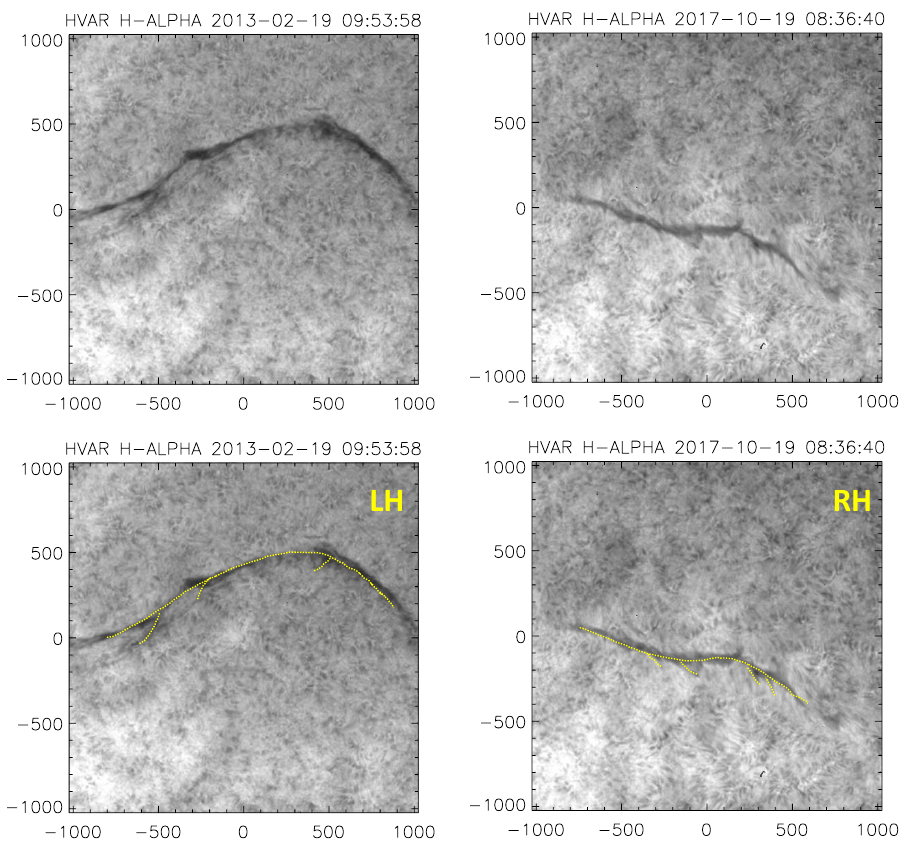}}
\caption{Examples of sinistral (left) and dextral (right) filaments from  the HVAR H$\alpha$ catalog with left-bearing and right-bearing barbs depicting left-handed and right-handed chirality, respectively. The spine and barbs are highlighted yellow in the bottom panels to guide the eye.}
\label{fig_chirality}
\end{figure}

\mateja{We next analyse the morphology of the filament, during its ``quiet'' phase (when it is not active) to obtain ``magnetic insight'' into the magnetic structure that holds them. Namely, depending on their orientation, filament channels can be dextral or sinistral, corresponding to the positive and negative helicity of the flux rope, respectively. A typical filament is characterised by the spine, legs, and barbs (small extensions from the spine), where the chirality is reflected in the observation of the filament barbs. If the barbs extend from the filament spine from left-to-right, the chirality is dextral (positive helicity, right-handed flux rope), whereas if the barbs extend right-to-left, the chirality is sinistral (negative helicity, left-handed flux rope) \citep{martin98, palmerio17, parenti14}. Examples of the two types of filaments are shown in Figure \ref{fig_chirality}. Note that these features are not equally well recognised in all filaments. Filaments form within filament channels, usually situated along polarity inversion lines (PILs). Not all filament channels are filled with filament plasma, though the observation of empty filament channels is rare \citep{parenti14}. However, sometimes only segments of a filament channel is filled with filament plasma, thus this might  hamper the observation of their properties in H$\alpha$.}

\mateja{Note that while some filaments never show any notable activity, some eventually start to oscillate or erupt. These eruptions are violent phenomena during which the prominence disappears and are often accompanied by solar flares and coronal mass ejections \citep[see e.g.][and references therein]{schmieder13}. For these active filaments we additionally aim to characterise type of activity.  We characterise the prominence/filament eruption as full (filament fully dissapears), partial (only part of the filament erupts), or failed (filament starts to erupt but fails). }The eruption is usually preceded by an activation phase, during which the prominence may exhibit mass motions in its spine or footpoints (e.g. oscillations), fragmentary brightening and fading (due to heating), EUV brightening (due to reconnections), deformation and helical rotation, or slow lift-off \citep{parenti14}. But this may not always be easily visible/recognisable in HVAR H$\alpha$ observations, therefore we do not characterise the activation phase. We note that we did try to recognize oscillations of prominences/filaments, which are observed either as oscillatory motions or winking in H$\alpha$ \citep[e.g.][]{vrsnak93}. Typical oscillation periods range from 32 to 110 minutes \citep{luna18}, and the velocity amplitudes are typically around tens of km/s. Thus, using daily observation videos which do not remove the atmospheric effects (shaking of the images) or FOV changes hamper observation of filament oscillation. While we do not exclude the possibility of observing and analysing possible prominence/filament oscillation in HVAR H$\alpha$ observations, it is beyond the scope of this study.

\begin{figure}
\centerline{\includegraphics[width=0.95\textwidth]{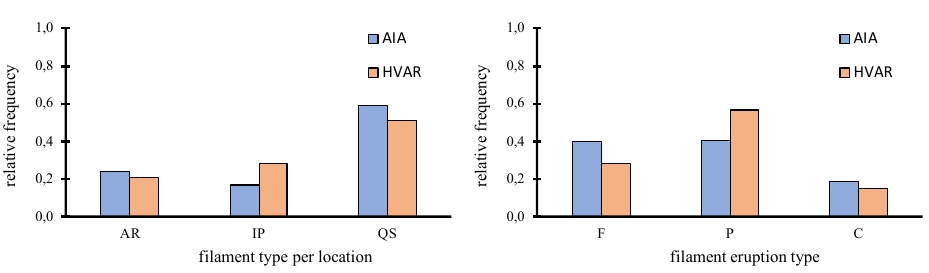}}
\caption{Left: Relative frequencies of observed filament type per location (AR=active region, IP=intermediate, QS=quiescent) for HVAR and AIA filament eruption catalogs; Right: Relative frequencies of observed eruption type (F=full, P=partial, C=confined) for HVAR and AIA filament eruption catalogs.}
\label{fig_filament_stat}
\end{figure}

\mateja{We catalog a total number of 276 filaments and prominences in 366 observations (some are observed multiple times). A large majority are filament observations (78.7\%) and most of the filament/prominence observations are quiescent (50.8\%), whereas we find 28.7\% observations of intermediate filaments/prominences and only 20.5\% from ARs. We are able to determine chirality in 96 filament observations, with 57.3\% of them estimated as right-handed. Finally, we observe eruption of the filament/prominence in 92 cases, out of which 28.3\% are observed as full eruptions, 56.5\% as partial eruptions and 15.2\% as failed (or confined) eruptions.} The catalog of HVAR filament/prominence observations is publicly available through figshare\footnote{\url{https://doi.org/10.6084/m9.figshare.23530416.v1}}.

\begin{figure}
\centerline{\includegraphics[width=0.95\textwidth]{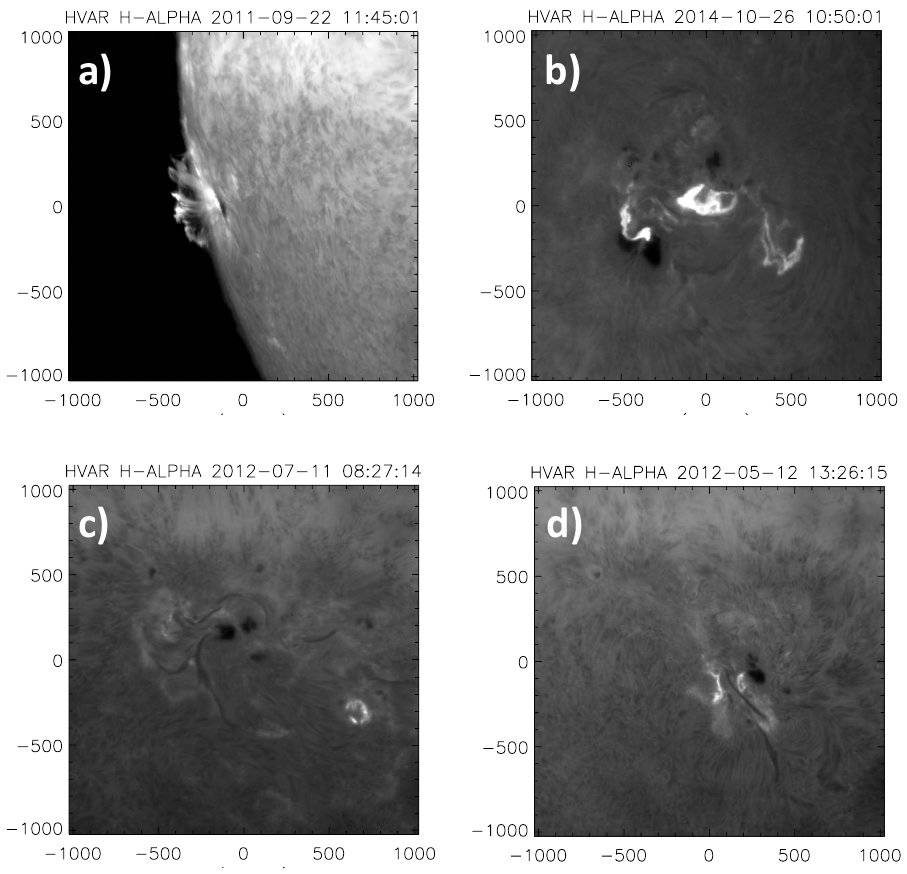}}
\caption{Examples of flare types in HVAR H$\alpha$ flare catalog: a) flare with post eruptive arcades; b) flare with asymmetric ribbons; c) flare with circular ribbon; d) symmetric two-flare ribbon.}
\label{fig_flare_type}
\end{figure}

\mateja{We compare list of HVAR H$\alpha$ eruptive prominences/filaments with AIA Filament Eruption Catalog\footnote{\url{https://aia.cfa.harvard.edu/filament/}} \citep{mccauley15}, which catalogs filament eruptions using full disc solar images in 171, 193, and 304 A wavelengths. In the HVAR observation time period AIA Filament Eruption Catalog detected 829 eruptive filaments (almost 10 times more). We compare the number of active region, intermediate and quiescent filaments in the two catalogs. The comparison is shown in the left panel of Figure \ref{fig_filament_stat}. We can see that there are no notable differences between the two catalogs. We also compare the number of full, partial and failed/confined eruptions in the two catalogs (right panel in Figure \ref{fig_filament_stat}), where we observe a slight bias in HVAR catalog towards observation of partial eruptions. This could be related to the fact that the observer is likely to choose to observe a filament which already showed some activity.}

Next we catalog prominent flares, where each flare is associated with an AR (NOAA AR number) and the GOES class via comparison with the entries in the Solar Monitor. We additionally categorise flares according to their observed eruptive signatures, i.e. whether their H$\alpha$ signatures include loops or ribbons. The loops are often referred to as post eruptive arcades (PEA) or post flare loops and are typically observed off the disc, whereas the ribbons are observed on disc. \mateja{Note that the ribbons are situated at the footpoints of both hot and cool loops, where the inner edge of the cool H$\alpha$ loops roughly correspond to the inner edge of the ribbons, whereas the outer edge of the ribbons is typically associated to hot X--ray loops, not observed here.} During the course of a flare, the separation between the ribbons increases, and the loops grow larger with time \citep[e.g.][]{priest02}. We additionally characterise flare according to the shape of their ribbons as symmetric two-ribbon flares, asymmetric, or circular-shaped. The four examples of flare types are shown in Figure \ref{fig_flare_type}.

\begin{figure}
\centerline{\includegraphics[width=0.95\textwidth]{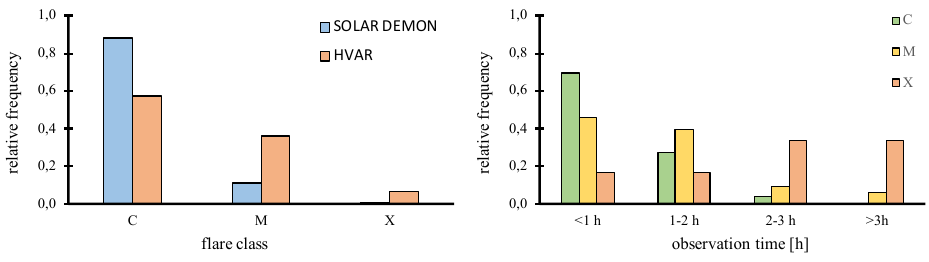}}
\caption{Left: Relative frequencies of observed flare classes in Solar demon flare catalog and HVAR catalog; Right: Relative frequencies of observing time dedicated to a flare for different flare classes in HVAR catalog.}
\label{fig_flare_stat}
\end{figure}

 The catalog contains a total of 91 prominent flares, most of which are GOES C-class flares (52), with 33 M-class flares and only 6 X-class flares. We additionally checked WL observations during 6 detected X-class flares searching for WL flares, but we found no WL signatures corresponding to the observed H$\alpha$ flares. We specifically double checked each GOES X-class flare observed by HVAR H$\alpha$ telescope in HVAR WL observation and HMI continuum observation. We found one WL flare observed by HMI continuum (barely visible) which is not observed by HVAR WL. In addition, one WL flare observed by HMI continuum was not seen in HVAR WL observation, because the start and peak of the flare was missing in the HVAR observations. We find that majority of flares are observed as ribbons on disc (86) with only 5 flares showing clear signatures of post-eruptive arcades. Regarding the ribbons, most flares are characterised as having asymmetric ribbons (71), with 8 flares appearing as circular-shaped ribbons and only 7 clear symmetric two-ribbon flares. The catalog is publicly available through figshare\footnote{\url{https://doi.org/10.6084/m9.figshare.23533755.v1}}.

\begin{figure}
\centerline{\includegraphics[width=0.95\textwidth]{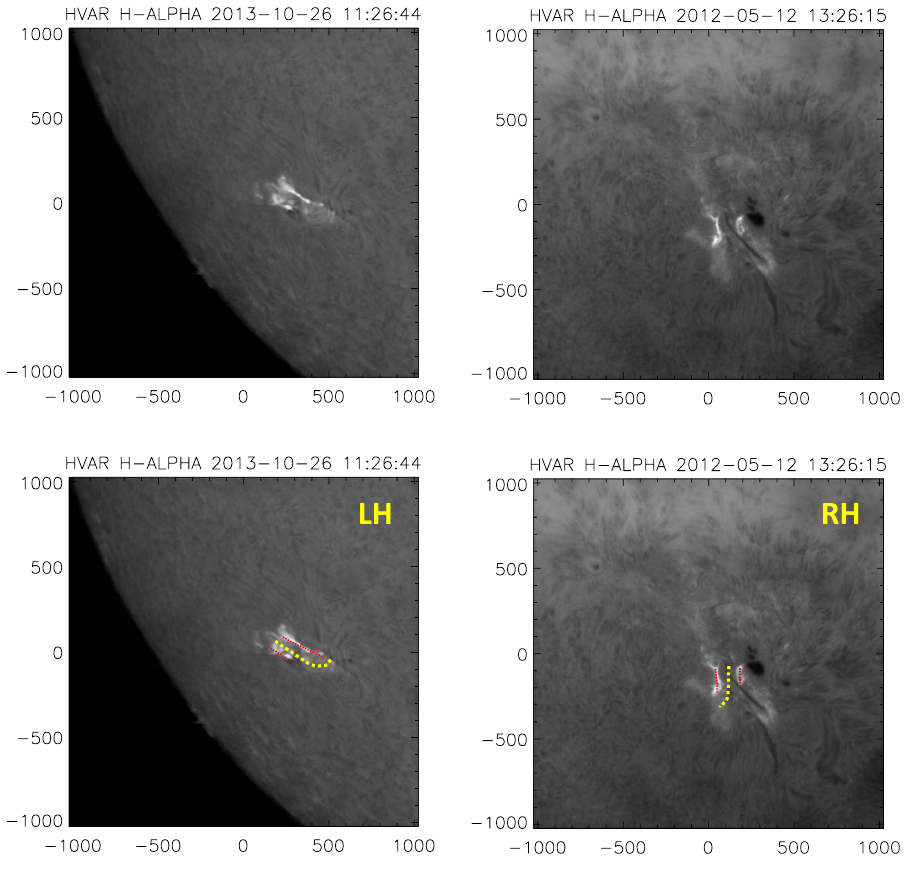}}
\caption{Examples of reverse J-shaped (left-handed) and J-shaped (right-handed) two flare ribbons are given in the left and right panels, respectively. The ribbons are highlighted red and the reverse J-shape and the J-shape, respectively, are highlighted yellow in the bottom panels to guide the eye.}
\label{fig_ribbons}
\end{figure}

\mateja{We compare list of HVAR H$\alpha$ flares with a catalog of flares detected in full disc EUV images, Solar Demon\footnote{\url{https://www.sidc.be/solardemon/}} \citep{kraaikamp15}. In the HVAR observation time period Solar demon detected 5183 flares of C-class or higher (4572 C-class flares, 567 M-class flares and 44 X-class flares). The comparison is shown in the left panel of Figure \ref{fig_flare_stat} where it can be seen that the relative frequency of M and X-class flares is higher in HVAR catalog, whereas the relative frequency of C-class flares is lower. In addition, we analyse how much observing time is dedicated to different types of flares in the HVAR catalog (see right panel in Figure \ref{fig_flare_stat}. We find that X-class flares tend to be observed longer than C and M-class flares. HVAR observations therefore seem biased towards more frequently observing stronger flares and observing them in longer time periods.}

Finally, we additionally analyse symmetric two-flare ribbons. The characteristics of ribbons can indicate the helicity of the corresponding eruptive structure (i.e. flux rope). A characteristic double J-shape and reverse J-shape of the two flare ribbons indicates positive and negative chirality, respectively \citep{demoulin96,palmerio17}. Another indication of the helicity of the corresponding eruptive structure can be the displacement of the ribbons along the PIL, which is related to the shear. \mateja{Assuming the PIL lies between the two ribbons, for positive chirality the left ribbon is displaced downwards and the right ribbon upwards} and vice versa for negative chirality \citep{demoulin96,palmerio17}. The examples of two distinct chiralities are given in Figure \ref{fig_ribbons} and all 7 symmetric two-flare ribbons with their properties are listed in Table \ref{tab_ribbons}.

\begin{table}
\caption{List of symmetric two-flare ribbons from the HVAR H$\alpha$ flare catalog with their estimated chirality.}
\label{tab_ribbons}
\centering
\small
\begin{tabular}{ccccccc} 
\hline
date	&	start time	&	end time	&	NOAA AR	&	GOES class	&	chirality	&	QI	\\
\hline
2011-08-02	&   05:46	&   07:41	&   11261	&   M1.4	&   R	&   3   \\
2012-05-12	&	13:20	&	14:37	&	11476	&	C2.5	&	R	&	2	\\
2013-06-19	&	07:29	&	08:23	&	11776	&	C3.5	&	L	&	1	\\
2013-10-26	&	10:55	&	14:48	&	11882	&	M1.8	&	L	&	3	\\
2014-07-06	&	08:08	&	08:58	&	12109	&	C2.9	&	L	&	2	\\
2014-07-25	&	06:58	&	07:42	&	12121	&	C2.2	&	L	&	2	\\
2015-07-02	&	05:58	&	07:15	&	12373	&	C1.4	&	R	&	2	\\
\hline
\end{tabular}
\end{table}

\section{Case study}
\label{case_study}

\begin{figure}
\centerline{\includegraphics[width=0.99\textwidth]{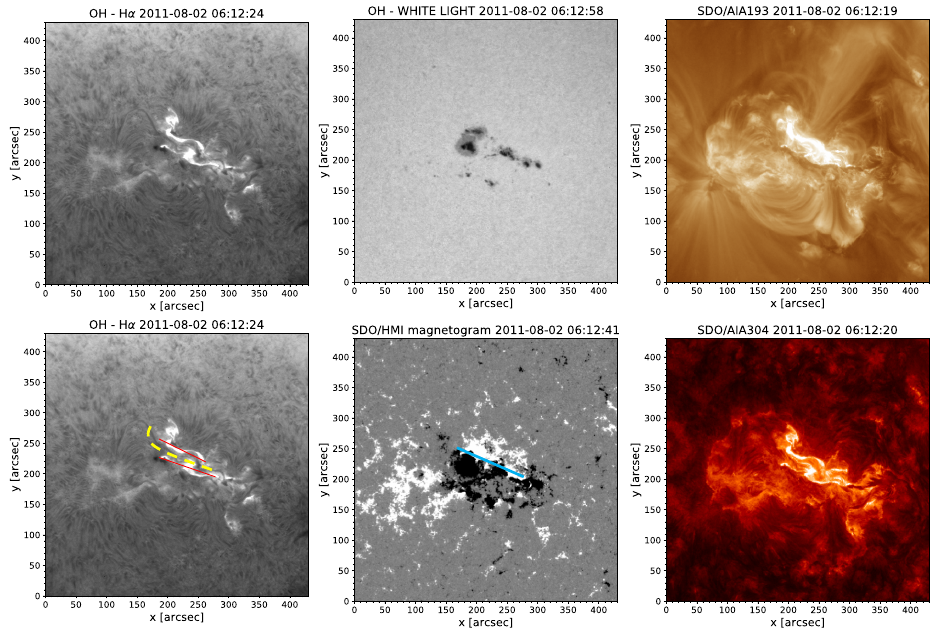}}
\caption{Flare and active region overview in HVAR H$\alpha$ and WL observations (upper left and middle panels, respectively), SDO/HMI magnetogram (lower middle panel) and SDO AIA 193 A and 304 A wavelengths (right upper and lower panels, respectively). HVAR H$\alpha$ figure is repeated in the lower left panel with ribbons outlined red and J-shape outlined yellow, to guide the eye. The rough estimate of the PIL orientation is highlighted with the blue line in SDO/HMI magnetogram figure.}
\label{fig_LCS}
\end{figure}

\mateja{We demonstrate the feasibility of the catalog on a case study of the flare detected on 2 August 2011 and related Sun-to-Earth phenomena. The event is listed in the HVAR H$\alpha$ flare catalog as a two-ribbon flare with start of observation 2 August 2011 05:46 and end of observation 07:41 UTC. The associated GOES X-ray class is M1.4 according to Solar Monitor and M2 according to Solar Demon. The flare is observed in AR11261, which, according to the HVAR WL catalog, is of Fkc and $\beta \gamma \delta$ classes according to the McIntosh and Mt Wilson classifications, respectively. At the time of the eruption the AR is located around N15W15. In the same AR a small filament erupts around 06:31 UTC south west of the flare ribbons, most likely triggered by instability caused by the first eruption. However, not being very prominent, this small filament eruption is not cataloged. According to the GOES soft X-ray light curve the start, peak and end of the flare are 05:24, 06:26, and 09:18 UTC, respectively.}

\mateja{The flare is shown in Figure \ref{fig_LCS}, where it can be seen that the ribbons show a characteristic J-shape, with the left ribbon displaced downwards and the right ribbon upwards. This indicates positive chirality, i.e. right handedness of the corresponding flux rope. Comparing the SDO/HMI magnetogram with HVAR H$\alpha$ image, we can see that the two ribbons are roughly aligned with the PIL, in the NE-SW direction. This indicates tilt between 0 and -45 degrees (measured from the solar equator in the counterclokwise direction). The ribbons are also visible in SDO/AIA 304, whereas in SDO/AIA 193 we see bright post eruptive arcades. However, the characteristic J-shape of the ribbons and their orientation seem slightly better visible in H$\alpha$ than EUV.}

\mateja{We next try to associate the flare to a coronal mass ejection using the temporal and spatial causality criteria from \citep{vrsnak05} applied to the  CDAW SOHO/LASCO CME catalog\footnote{\url{https://cdaw.gsfc.nasa.gov/CME_list/}}. According to these criteria, a flare is most likely physically related to the CME if its start time is within $\pm 1$h of the CME start time (derived by extrapolating CME kinematics to solar surface) and if its position angle, measured counterclockwise from north, is within the opening angle of the CME. The flare position angle is $307^{\circ}$ and we find it temporally and spatially related to the partial halo CME with first C2 appearance 2011-08-02 06:36 UTC, which has a position angle of $288^{\circ}$ and opening angle of $268^{\circ}$. The CME is also observed in STEREO-A and -B SECCHI/COR2 coronagraphs, thus providing us with observations from multiple vantage points.} 

\begin{figure}
\centerline{\includegraphics[width=0.99\textwidth]{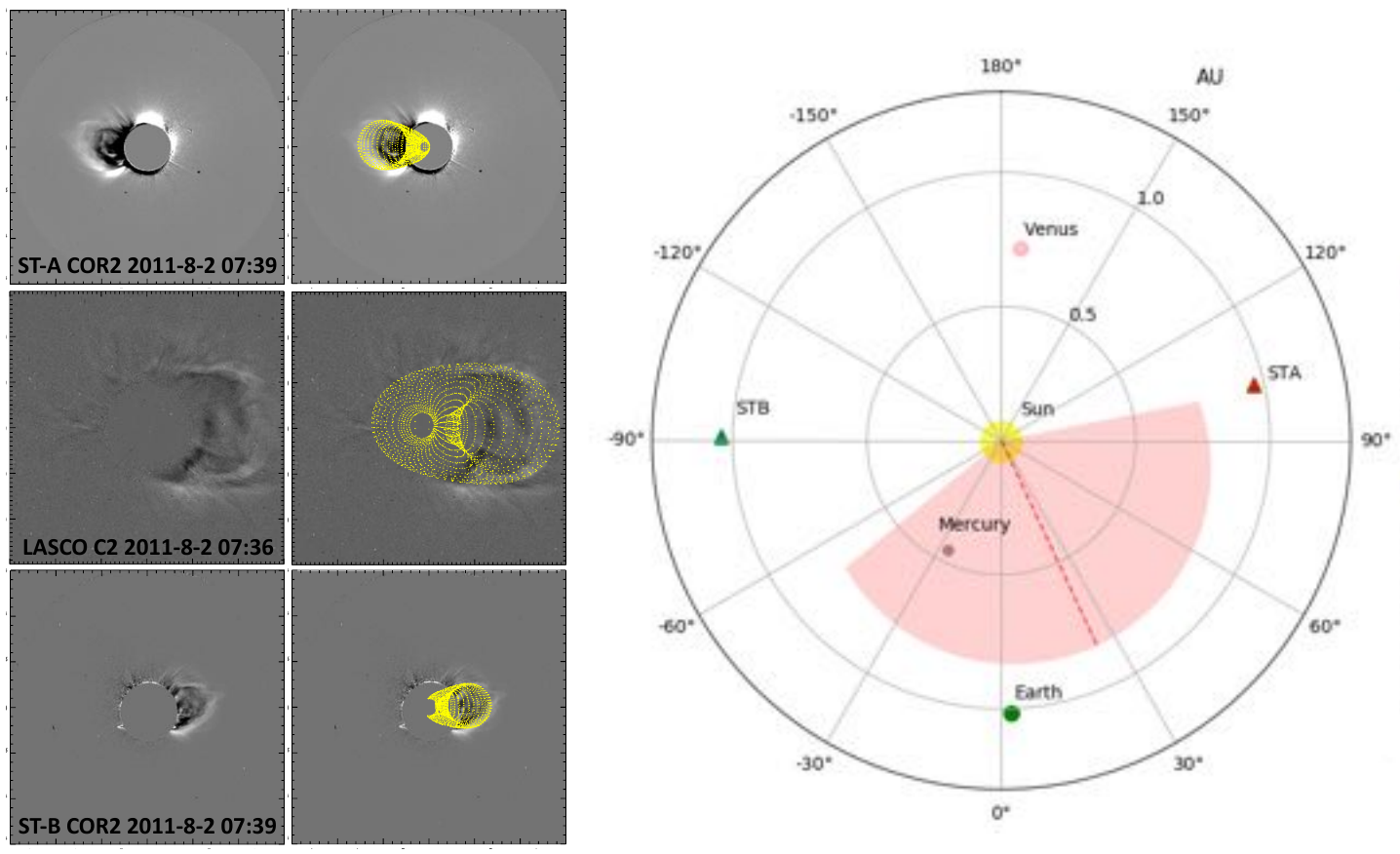}}
\caption{Left: CME observation in coronagraphs from three different vantage points and the corresponding 3D reconstruction of the CME using the GCS model (longitude=$26^{\circ}$, latitude=$9^{\circ}$, tilt=$2^{\circ}$, height=8$\mathrm{R_{\odot}}$, ratio=0.49, halfangle=$44^{\circ}$). Right: Constellation of spacecraft and the direction and extent of the CME (taken from the DBEMv3 tool, for details see main text).}
\label{fig_CME}
\end{figure}

\mateja{We perform a 3D reconstruction of the CME geometry using the Graduated Cylindrical Shell model \citep[GCS,][]{thernisien06}. GCS model assumes CME has a geometry of a croissant, with conical legs centered at the Sun, connected by a toroid of varying cross-section (largest at apex, smallest at the legs) and a pseudo-circular front. The CME geometry and position is fully defined in GCS by six parameters: longitude of the apex, latitude of the apex, tilt of the croissant axis (with respect to solar equator), height of the apex, aspect ratio (i.e. opening angle of the legs) and half angle (full opening angle of the croissant, measured from outer edges of the legs). We perform the GCS reconstruction by manually fitting projections of the croissant geometry to images from three vantage points (STEREO-A/COR2, SOHO/LASCO/C2, and STEREO-B/COR2) at the time when  CME is best observed in all three spacecraft. Note that there is a slight difference in the timing of the CME measurements between three spacecraft, because STEREO and SOHO do not have identical time resolution. The difference of 3 minutes however, does not significantly influence the results, as the CME is not very fast ($v=712\,\mathrm{kms}^{-1}$, according to the SOHO/LASCO catalog). The CME longitude and latitude (see Figure \ref{fig_CME}) correspond well with the position of the AR and the CME tilt corresponds well with the orientation of the PIL (both low inclined).} 

\begin{figure}
\centerline{\includegraphics[width=0.99\textwidth]{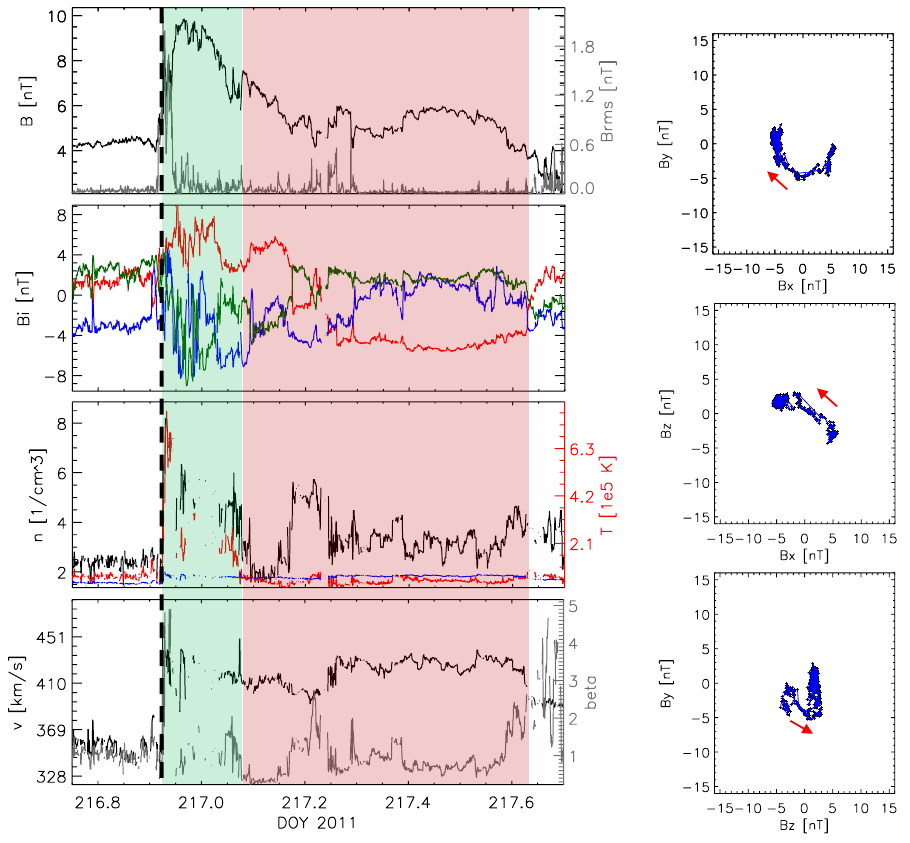}}
\caption{Left: in situ measurements of the ICME that arrived at Earth 4 August 2011. The panels show (top to bottom) total magnetic field and magnetic field fluctuations (root-mean-square, RMS); GSE magnetic field components x (red), y (blue) and z (green); plasma density, temperature and expected temperature calculated according to \citet{lopez87}; and plasma flow speed and plasma beta. The dashed line marks the shock, sheath is shaded green, whereas the magnetic ejecta is shaded red. Right: Hodograms depicting the behaviour of the magnetic field within borders of the magnetic ejecta in different planes. The red arrow marks the direction of rotation.}
\label{fig_insitu}
\end{figure}

\mateja{Next, we search for possible interplanetary counterpart of the CME (ICME). For that purpose, we run the Drag-based ensemble model tool \citep[DBEMv3 tool,][]{calogovic21} from the ESA Heliospheric Weather Expert Service Centre\footnote{\url{https://swe.ssa.esa.int/heliospheric-weather}}. This is an ensemble version of the drag-based model \citep[DBM][]{vrsnak13}, which takes into account input uncertainties and accordingly provides uncertainty intervals for the output. As input we use parameters obtained from the GCS reconstruction, supplemented with initial CME kinematics from SOHO/LASCO catalog (extrapolated to $20\,_{\odot}$, the inner boundary of the model). The DBEMv3 input used is: start time = 11:30 $\pm$ 30 min; drag parameter = 0.2 $\pm$ 0.1 $10^{-7}\mathrm{km}^{-1}$; solar wind speed = 350 $\pm 50\,\mathrm{kms^{-1}}$ (estimated based on the observation of in situ solar wind speed near Earth around the time of the eruption); solar wind speed = 712 $\pm 50\,\mathrm{kms^{-1}}$; CME angular half-width = 75.8 $\pm 10^{\circ}$ (derived from GCS reconstruction parameters by the tool) and longitude = 26 $\pm 10^{\circ}$. All input uncertainties were selected as suggested by the DBEMv3 tool (default uncertainty values).  DBEM forecasts the arrival of the ICME at Earth between 5 August 2011 03:15 and 13:50 UTC, with the arrival speed between 435 and 525 $\mathrm{kms^{-1}}$. We do indeed find an ICME within the predicted time period. We check \citet{richardson24} ICME catalog and it has an entry of an ICME where the shock arrival is 4 August at 21:53 UTC and the start of the magnetic ejecta is 5 August 2011 05:00 UTC. Moreover, they associate this ICME to the same CME. The average ICME speed is 430 $\mathrm{kms^{-1}}$, which is only slightly lower than predicted by DBEMv3.}

\mateja{Finally, we analyse in situ measurements of the ICME, shown in Figure \ref{fig_insitu}. We use the 1 minute plasma and magnetic field data in the Geocentric solar ecliptic (GSE) system provided by the OMNIWeb database \citep{king05}. Shock arrival is observed 4 August 2011 around 22:00 UTC (DOY 216.9), followed by the highly turbulent compressed and heated plasma of the ICME sheath region. After the sheath, the magnetic ejecta starts 5 August 2011 around 03:00 UTC (DOY 217.15), which is observed as the rotating magnetic field and low plasma temperature, though other parameters do not show typical characteristics \citep[for overview on ICME sheath and magnetic ejecta characteristics see e.g.][]{kilpua17}. We can see in the in situ measurements and more so in hodograms in Figure \ref{fig_insitu} that throughout the magnetic ejecta all three components show rotation. Based on the behaviour of the rotation, it is possible to deduce the orientation and chirality of the flux rope \citep[see e.g.][]{nieves-chinchilla19}.}

\begin{figure}
\centerline{\includegraphics[width=0.99\textwidth]{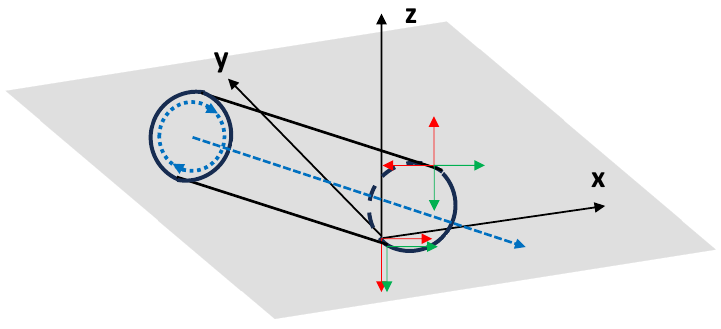}}
\caption{A sketch visualising the flux rope passing at some oblique angle $\Phi$ in xy plane and $\Theta$ in yz plane in GSE coordinates. The blue dotted circle with an arrow marks the direction of the poloidal field, whereas the long blue dotted arrow marks the direction of the axial field. Red and green arrows at the leading/trailing edge of the flux rope mark components of the poloidal and axial field, respectively (in the $B_z$ and $B_y$ direction).}
\label{fig_sketch}
\end{figure}

\mateja{The $B_x$ shows rotation from positive to negative values, indicating that the flux rope axis is not vertical to the radial heliocentric direction. $B_z$ and $B_y$ both show partial rotation, from negative values to $\approx 0$. This indicates that the flux rope axis is not aligned neither with y- or z-axis and that at the leading edge of the flux rope the poloidal magnetic field is oriented south-west. Consequently, we would expect the poloidal field at the back of the flux rope to be oriented north-south. As  $B_z$ and $B_y$ do not undergo full rotation, this could indicate that the axial magnetic field counters the poloidal axial field in $B_z$ and $B_y$ directions at the back of the flux rope, i.e. that it points to the south-west. This is a configuration of a right-handed flux rope of positive chirality, in agreement with the chirality estimated based on the flare ribbons. A sketch visualising this is shown in Figure \ref{fig_sketch}.}

\section{Summary and Conclusions}
\label{summary}

We compile the catalog of Hvar Observatory solar observations in the time period corresponding to regular digitally stored chromospheric and photospheric observations 2010-2019. \mateja{We make basic characterisation of observed phenomena and compare them to catalogs which are based on full disc solar images.} 

Given its resolution and FOV, HVAR WL observations are suitable to be used for study of photospheric morphology and activity related to ARs and WL flares. We thus compile a catalog of observed ARs consisting of 1100 entries, where each AR is classified according to McIntosh and Mt Wilson classifications. We search for WL flares in HVAR WL observations for the given time period, but we do not find any. \mateja{The comaprison of HVAR AR catalog with catalogs based on full disc solar observations reveals that HVAR observations are biased towards towards more frequently observing more complex ARs and observing them in longer time periods. This is most likely related to the fact that HVAR telescopes have small FOV that does not cover the whole solar disc. Therefore, the observer has to choose what part of the Sun to observe and is likely to choose a place where they expect more activity.}

Regarding the  HVAR H$\alpha$ observations, we find that the most suitable phenomena to be observed and analysed are prominences/filaments and flares. There is also the possibility of observing Moreton waves, but in the given time span, we do not observe any. We thus focus on cataloging flares and prominences/filaments. In order for the catalog to hold more information than simply observation time, we only catalog conspicuous phenomena, which we can characterise. We thus compile a catalog of prominences/filaments which we characterise according to their location, chirality (if possible) and eruptive signatures. \mateja{The catalog contains a total number of 276 filaments and prominences in 366 observations (some are observed multiple times), where we are able to determine chirality in 96 filament observations (26.2\%) and find a total of 92 eruptive filaments (25.1\%). The comparison of the eruptive filaments with the AIA eruptive filament catalog reveals that there is a slight bias in HVAR catalog towards observation of partial eruptions. This could be related to the fact that the observer is likely to choose to observe a filament which already showed some activity.}

In the flare catalog we focus on their observed eruptive signatures (loops or ribbons) and their shape. \mateja{Out of 91 cataloged flares majority are observed as ribbons on disc (86, i.e. 94.5\%) with only 5 flares (5.5\%)showing clear signatures of post-eruptive arcades. Regarding the ribbons, most flares are characterised as having asymmetric ribbons (71), with 8 flares appearing as circular-shaped ribbons and only 7 clear symmetric two-ribbon flares, which we additionally analyse to determine the chirality of the associated erupting structure. the comparison of the flare catalog with a catalog of flares detected using full disc solar images reveals that HVAR observations seem biased towards more frequently observing stronger flares and observing them in longer time periods.}

\mateja{Finally, we demonstrate the feasibility of the catalog on a case study of the flare detected on 2 August 2011 and related Sun-to-Earth phenomena. The analysis of the flare characteristics observed in HVAR H$\alpha$ observations reveals a right-handed structure of low inclination. We associate the flare to a CME and perform 3D reconstruction, which is in agreement with the low coronal signatures. We next associate an ICME to the CME, where we find that the magnetic structure of the ICME shows inclination and chirality in good agreement with the solar observation.}

We note that the movies of the daily observations used for this study are publicly available at Hvar Observatory webpage\footnote{\url{http://oh.geof.unizg.hr}}, the data (in FITS or jpeg format) is available upon request and the catalogs are publicly available via figshare\footnote{\url{https://figshare.com/s/da38ce807968ecf92bd3}}\footnote{\url{https://figshare.com/s/24b2f98a4fbc271aab5d}}\footnote{\url{https://figshare.com/s/85b87d210b12a1d4c0f6} (NOTE: these are private figshare links, public ones with DOI will be added once the paper is accepted and the list is published in figshare.)}

\begin{acks}
This research has received financial support from the European Union’s Horizon 2020 research and innovation program under grant agreement No. 824135 (SOLARNET). M.D. acknowledges the support by the Croatian Science Foundation under the project IP-2020-02-9893 (ICOHOSS).
We thank the team that maintains the CDAW SOHO/LASCO CME catalog. This CME catalog is generated and maintained at the CDAW Data Center by NASA and The Catholic University of America in cooperation with the Naval Research Laboratory. SOHO is a project of international cooperation between ESA and NASA. We thank the science teams of SDO, SOHO, STEREO, ACE and Wind for providing the data and maintaining the spacecraft. This study has made use of the JHelioviewer software provided by ESA.
We thank Hvar Observatory technicians, Nik\v{s}a Novak and Toni Viskovi\'c for performing the regular observations with the solar telescope. Our special thanks goes to late Vlado Ru\v{z}djak, who for years coordinated solar observations at Hvar Observatory. In addition we thank W. P\"otzi from Kanzelh\"ohe Observatory who is persistently aiding in maintenance of the telescopes.
\end{acks}

\bibliographystyle{spr-mp-sola}
\bibliography{REFs}  

\end{article} 

\end{document}